\documentclass[hyper]{JHEP} 

\usepackage{epsfig}

\newcommand\fverb{\setbox\pippobox=\hbox\bgroup\verb}

\newcommand\fverbdo{\egroup\medskip\noindent%

            \fbox{\unhbox\pippobox}\ }

\newcommand\fverbit{\egroup\item[\fbox{\unhbox\pippobox}]}

\newbox\pippobox

\title{Hamiltonian Formalism of General  Bimetric  Gravity}
\author{J. Kluso\v{n}\\
Department of
Theoretical Physics and Astrophysics\\
Faculty of Science, Masaryk University\\
Kotl\'{a}\v{r}sk\'{a} 2, 611 37, Brno\\
Czech Republic\\
E-mail: \email{klu@physics.muni.cz}} \preprint{}

 \abstract{We perform the Hamiltonian analysis of
 general bimetric gravity. We determine four first class constraints
 that are generators of the diagonal diffeomorphism. We
further analyze the  remaining constraints and we present an
evidence that these constraints should be the second class
constraints in order to have theory with the Hamiltonian constraint
as the first class constraint.  We also discuss the case of the
non-linear bimetric gravity and argue that it is very difficult to
eliminate the ghost mode. } \keywords{Massive Gravity}

\def\mR{\mathcal{R}}

\def\hf{\hat{f}}
\def\tK{\tilde{K}}

\def\bA{\mathbf{A}}
\def\bF{\mathbf{F}}
\def\tP{\tilde{P}}
\def\bV{\mathbf{V}}
\def\bW{\mathbf{W}}
\def\bN{\bar{N}}

\def\bB{\mathbf{B}}

\def\bN{\bar{N}}

\def\bE{\mathbf{E}}

\def\tu{\tilde{u}}
\def\tv{\tilde{v}}

\def\tnabla{\tilde{\nabla}}

\def\tmG{\tilde{\mG}}

\def\be{\begin{equation}}

\def\ee{\end{equation}}

\def\bea{\begin{eqnarray}}

\def\eea{\end{eqnarray}}

\def\tr{\mathrm{tr}\, }

\def\tmR{\tilde{\mR}}

\def\bE{\mathbf{E}}
\def\bX{\mathbf{X}}

\def\bR{\bar{R}}

\def\bz{\mathbf{z}}

\def\tr{\mathrm{Tr}}

\def\bG{\mathbf{G}}

\def\bx{\mathbf{x}}
\def\by{\mathbf{y}}

\newcommand{\hg}{\hat{g}}

\newcommand{\mG}{\mathcal{G}}

\def\mV{\mathcal{V}}

\def \bA{\mathbf{A}}

\newcommand{\bT}{\mathbf{T}}
\def\bmR{\bar{\mR}}

\def\pb #1{\left\{#1\right\}}

\begin{document}
\section{Introduction}
The bimetric theory of gravity  were introduced in
\cite{Isham:1971gm} in order to describe the interaction of gravity
with a massive spin $2-$meson. These theories were also investigated
recently due to their cosmological solutions
\cite{Hassan:2012rq,Hassan:2012wr,Nomura:2012xr,Baccetti:2012re,Baccetti:2012bk,
Banados:2009it,Banados:2008fi,Blas:2005yk,Damour:2002ws}. Generally,
bimetric theories contain  one massive and one massless spin$-2$
field and one scalar ghost mode. As a result the bimetric theories
of gravity are plagued by the same problems as the general massive
theories of gravity.

On the other hand recent formulation of the massive gravity known as
the non-linear massive gravity seems to be free of ghosts
\cite{deRham:2010ik,deRham:2010kj}, for further improvement, see
\cite{Hassan:2011hr,Hassan:2011vm}. This theory was further extended
in \cite{Hassan:2011tf} where the theory was formulated with general
reference metric. The Hamiltonian analysis of given theory was
performed  in  several papers
\cite{Kluson:2011qe,Kluson:2011aq,Golovnev:2011aa,Kluson:2011rt,Kluson:2012gz,Kluson:2012wf,Kluson:2012zz}
with the most important results derived in
\cite{Hassan:2011ea,Hassan:2012qv}  with the outcome that this
non-linear massive theory possesses one additional constraint and
the resulting constraint structure is sufficient for the elimination
of the ghost degree of freedom.

Due to the fact that the non-linear massive gravity with general
reference metric is free of ghost it is tempting to generalize this
theory by introducing the kinetic term for the reference metric
which now becomes dynamical. As a result  $\hg_{\mu\nu}$ and
$\hf_{\mu\nu}$ come in the symmetric way in the action which means
that the non-linear massive gravity was generalized to the bimetric
theory of gravity. This important step was performed in
\cite{Hassan:2011zd}. Then it was argued in \cite{Hassan:2011ea}
 that the resulting
theory is the ghost free formulation of the bimetric theory of
gravity using the very detailed proof of the absence of ghosts in
the non-linear massive gravity performed here. However the question
of the absence of the ghosts in the new formulation of the bimetric
theory of gravity was questioned recently in \cite{Kluson:2013cy}
\footnote{This issue was also addressed in the recent paper
\cite{Soloviev:2013mia}.}. This analysis was based on the careful
Hamiltonian analysis of the bimetric gravity with redefined shift
function \cite{Hassan:2011zd}. We have argued that due to the fact
that the lapse function of the metric $\hf_{\mu\nu}$ is now Lagrange
multiplier whose value is determined by the dynamics of the theory
 it is not possible to find another additional constraint so
that the ghost mode is still present.
 On the other hand we showed that
there are no first class constraints corresponding to the generators
of the diagonal diffeomorphism and we argued that this is a
consequence of the redefinition of the shift function performed in
\cite{Hassan:2011zd}.

The goal of this paper is to extend the analysis presented in
 \cite{Kluson:2012ps} to the case of the
  Hamiltonian analysis of the bimetric theory of the
gravity with general potential between two metric $\hg_{\mu\nu}$ and
$\hf_{\mu\nu}$. We identify four the first class constraints
corresponding to the diagonal diffeomorphism and determine the
Poisson brackets among them in order to show that they obey the
right form of the algebra.  These results immediately solve the
issue found in \cite{Kluson:2013cy} since we are now able to
identify the four first class constraints that generate the diagonal
diffeomorphism in case of non-linear bimetric gravity.  As the next
step in our analysis we try to answer the question whether it is
possible to find an additional constraint in bimetric gravity which
could eliminate the ghost mode. In other words we analyze the time
development of the  four remaining constraints.  We present some
evidence that it is very difficult to have such a form of the
potential that will lead to the emergence of the additional
constraint.
 Then we analyze the square root form of the potential
and we argue that it is difficult to find an additional constraint
which suggests that the ghost mode is still presented in given
theory which also confirms the observation presented in
\cite{Kluson:2013cy}.

The structure of this paper is as follows. In the next section
(\ref{second}) we introduce the bimetric theory of gravity and find
its Hamiltonian formulation. Then in section (\ref{third}) we
calculate the Poisson brackets between constraints and find the
constraints that are generators of the diagonal diffeomorphism. In
section (\ref{fourth}) we analyze the time evolution of the  second
class constraints. Finally in conclusion (\ref{fifth}) we outline
our result and suggests possible extension of this work.
\section{Hamiltonian Formulation  of   Bigravity with General Potential}
\label{second}
 Let us consider  bimetric  theory of gravity
defined by the action
\begin{equation}\label{bigraction}
S=M_L^2\int d^4x \sqrt{-\hg}R(\hg)+M_R^2 \int d^4x\sqrt{-\hf}
R(\hf)-\mu\int d^4x (\det \hg\det\hf)^{1/4} \mV(H^\mu_{ \ \nu}) \ ,
\end{equation}
where we presume that the potential term is the general function of
\begin{equation}
H^\mu_{ \ \nu}=\hg^{\mu\rho}\hf_{\rho\nu} \ .
\end{equation}
When we require that the action  (\ref{bigraction}) is invariant
under following diffeomorphism transformations
\begin{equation}\label{diffov}
\hg'_{\mu\nu}(x')=
 \hg_{\rho\sigma}(x) \frac{\partial x^\rho}{\partial x'^\mu}
\frac{\partial x^\sigma}{\partial x'^\nu}
 \ ,
 \quad
\hf'_{\mu\nu}(x')= \hf_{\rho\sigma}(x) \frac{\partial
x^\rho}{\partial x'^\mu} \frac{\partial x^\sigma}{\partial x'^\nu} \
\end{equation}
we have to demand that  the potential term has to be functions of various
powers of $H^\mu_{ \ \nu}$ and their traces.
The goal of this paper is to extend the Hamiltonian analysis of the
particular bimetric gravity performed in \cite{Kluson:2012ps}
 to the case of the
general form of the potential $\mV$.

To begin with we introduce the $3+1$ decomposition of the four
dimensional metric $\hat{g}_{\mu\nu}$
\cite{Gourgoulhon:2007ue,Arnowitt:1962hi}
\begin{eqnarray}
\hat{g}_{00}=-N^2+N_i g^{ij}N_j \ , \quad \hat{g}_{0i}=N_i \ , \quad
\hat{g}_{ij}=g_{ij} \ ,
\nonumber \\
\hat{g}^{00}=-\frac{1}{N^2} \ , \quad \hat{g}^{0i}=\frac{N^i}{N^2} \
, \quad \hat{g}^{ij}=g^{ij}-\frac{N^i N^j}{N^2} \
\nonumber \\
\end{eqnarray}
together with the metric $\hf_{\mu\nu}$
\begin{eqnarray}
\hf_{00}&=&-M^2+L_i f^{ij}L_j \ , \quad \hf_{0i}=L_i \ , \quad
\hf_{ij}=f_{ij} \ , \nonumber \\
\hf^{00}&=&-\frac{1}{M^2} \ , \quad  \hf^{0i}=\frac{L^i}{M^2} \ ,
\quad \hf^{ij}=
f^{ij}-\frac{L^i L^j}{M^2} \ , \quad  L^i=L_jf^{ji} \ . \nonumber \\
\end{eqnarray}
To proceed further we use the well known relation \footnote{We
ignore the boundary terms.}
\begin{eqnarray}\label{Rfour}
{}^{(4)}R[\hg]&=&K_{ij}\mG^{ijkl}K_{kl}+R^{(g)} \ , \nonumber \\
{}^{(4)}R[\hf]&=&\tK_{ij}\tmG^{ijkl}\tK_{kl}+R^{(f)} \ , \nonumber \\
\end{eqnarray}
where $R^{(g)}$ and $R^{(f)}$ are three dimensional scalar
curvatures evaluated using the spatial metric $g_{ij}$ and $f_{ij}$
respectively and  where the extrinsic curvatures $K_{ij}$ and
$\tK_{ij}$ are defined as
\begin{equation}
K_{ij}=\frac{1}{2N}(\partial_t g_{ij}- \nabla_i N_j-\nabla_j N_i)\ ,
\quad
 \tK_{ij}=\frac{1}{2M}(\partial_t f_{ij}- \tnabla_i
L_j-\tnabla_j L_i) \ ,
\end{equation}
and where $\nabla_i$ and $\tnabla_i$ are covariant derivatives
evaluated using the metric components $g_{ij}$ and $f_{ij}$
respectively. Finally note that $\mG^{ijkl}$ and $\tmG^{ijkl}$ are
de Witt metrics defined as
\begin{equation}
\mG^{ijkl}=\frac{1}{2}(g^{ik}g^{jl}+g^{il}g^{jk})-g^{ij}g^{kl} \ ,
\quad  \tmG^{ijkl}=\frac{1}{2}(f^{ik}f^{jl}+f^{il}f^{jk})-
f^{ij}f^{kl} \
\end{equation}
with inverse
\begin{equation}
\mG_{ijkl}=\frac{1}{2}(g_{ik}g_{jl}+ g_{il}g_{jk})-\frac{1}{2}
g_{ij}g_{kl} \ , \quad \tmG_{ijkl}=\frac{1}{2}(f_{ik}f_{jl}+
f_{il}f_{jk})-\frac{1}{2} f_{ij}f_{kl} \
\end{equation}
that obey the relation
\begin{equation}
\mG_{ijkl}\mG^{klmn}=\frac{1}{2}(\delta_i^m\delta_j^n+
\delta_i^n\delta_j^m)  \ , \quad
\tmG_{ijkl}\tmG^{klmn}=\frac{1}{2}(\delta_i^m\delta_j^n+
\delta_i^n\delta_j^m)  \ .
\end{equation}
Using (\ref{Rfour}) we rewrite the action (\ref{bigraction}) into
the form that is suitable for the Hamiltonian analysis
\begin{eqnarray}\label{SFRbi}
S&=&\int dt L=M_g^2 \int d^3\bx dt
\sqrt{g}N[K_{ij}\mG^{ijkl}K_{kl}+R^{(g)}
]+ \nonumber \\
&+& M_f^2\int d^3 \bx dt \sqrt{f}M [\tK_{ij}\tmG^{ijkl}\tK_{kl}+
R^{(f)}]-\mu \int d^3\bx dt g^{1/4}f^{1/4}\sqrt{NM}\mV \ .
 \nonumber \\
\end{eqnarray}
Then  from (\ref{SFRbi}) we determine following conjugate momenta
\begin{eqnarray}
\pi^{ij}&=&\frac{\delta L}{\delta \partial_t g_{ij}}=
M_g^2\mG^{ijkl}K_{kl} \ , \quad \rho^{ij}= \frac{\delta L}{\delta
\partial_t f_{ij}}= M_f^2\tmG^{ijkl}\tK_{kl} \ ,
\nonumber \\
\pi_i&=&\frac{\delta L}{\delta \partial_t N^i}\approx 0 \ , \quad
\rho_i=\frac{\delta L}{\delta \partial_t L^i}\approx 0 \ , \nonumber
\\
\pi_N&=&\frac{\delta L}{\delta \partial_t N}\approx 0 \ , \quad
\rho_M=\frac{\delta L}{\delta \partial_t M}\approx 0 \  \nonumber
\\
\end{eqnarray}
and then using the  standard procedure we derive following
Hamiltonian
\begin{eqnarray}
H=\int d^3\bx (N \mR_0^{(g)}+M \mR_0^{(f)}+ N^i\mR_i^{(g)}+
L^i\mR_i^{(f)}+
\mu \sqrt{NM}g^{1/4}f^{1/4}\mV) \  ,  \nonumber \\
\end{eqnarray}
where
\begin{eqnarray}
\mR_0^{(g)}&=&\frac{1}{M_g^2\sqrt{g}}
\pi^{ij}\mG_{ijkl}\pi^{kl}-M_g^2\sqrt{g}R^{(g)} \ , \quad
\mR_0^{(f)}= \frac{1}{M_f^2
\sqrt{f}}\rho^{ij}\tmG_{ijkl}\rho^{kl}-M_f^2\sqrt{f}R^{(f)} \ , \nonumber \\
\mR_i^{(g)}&=&-2g_{ij}\nabla_k\pi^{kj} \ , \quad \mR_i^{(f)}=
-2f_{ij}\tnabla_k\rho^{kj} \ . \nonumber \\
\end{eqnarray}
An important point is to identify four constraints that are generators  of the
 diagonal diffeomorphism. In order
to do this we proceed as in \cite{Damour:2002ws} and introduce
following variables
\begin{eqnarray}\label{defnewN}
\bN&=&\sqrt{NM} \ , \quad  n=\sqrt{\frac{N}{M}} \ , \quad
\bN^i=\frac{1}{2}
(N^i+L^i) \ , \quad  n^i=\frac{N^i-L^i}{\sqrt{NM}} \ , \nonumber \\
N&=&\bN n \ , \quad  M=\frac{\bN}{n} \ , \quad
L^i=\bN^i-\frac{1}{2}n^i\bN \ , \quad  N^i=\bN^i+\frac{1}{2}n^i\bN
 \ , \nonumber \\
\end{eqnarray}
where again clearly their conjugate momenta are the primary
constraints of the theory
\begin{equation}\label{Pbnprim}
P_{\bN}\approx 0 \ , \quad  p_n\approx 0 \ , \quad  P_i\approx 0 \ ,
\quad  p_i\approx 0 \ .
\end{equation}
Note that the canonical variables have following  non-zero Poisson
brackets
\begin{eqnarray}\label{canpb}
\pb{g_{ij}(\bx),\pi^{kl}(\by)}&=& \frac{1}{2}(\delta_i^k\delta_j^l+
\delta_j^k\delta_i^l)\delta(\bx-\by) \ , \quad
\pb{f_{ij}(\bx),\rho^{kl}(\by)}= \frac{1}{2}(\delta_i^k\delta_j^l+
\delta_j^k\delta_i^l)\delta(\bx-\by) \ , \nonumber \\
\pb{\bN(\bx),P_{\bN}(\by)}&=&\delta(\bx-\by) \ , \quad
 \pb{n(\bx),P_{n}(\by)}=\delta(\bx-\by) \ , \nonumber \\
\pb{\bN^i(\bx),P_j(\by)}&=&\delta^i_j\delta(\bx-\by) \ , \quad
\pb{n^i(\bx),p_j(\by)}=\delta^i_{j}\delta(\bx-\by) \ .
\nonumber \\
\end{eqnarray}
Further, using (\ref{defnewN}) we find following form of the matrix
 $H^\mu_{ \
\nu}$
\begin{eqnarray}
H^0_{ \ 0}&=& a+\frac{\bN^i}{\bN}v_i \ , \quad  H^0_{ \ j}=
\frac{1}{\bN}v_j \
, \nonumber \\
H^i_{ \ 0}
&=&-\bN^i a-\frac{\bN^i}{\bN}v_k\bN^k +a^i_{ \ j}\bN^j+\bN w^i
 \ ,\nonumber \\
  H^i_{ \ j}&=&a^i_{ \
j}-\frac{1}{\bN}\bN^iv_j \ ,
\nonumber \\
\end{eqnarray}
where
\begin{eqnarray}
v_i&=&\frac{f_{ij}n^j}{n^2} \ , \quad
w^i=\frac{1}{4n^2}n^i(n^mf_{mn}n^n)
-\frac{n^i}{2n^4}-\frac{1}{2}g^{im}f_{mk}n^k \ , \nonumber \\
a&=&\frac{1}{n^4}-\frac{n^if_{ij}n^j}{2n^2}
 \ , \quad
 a^i_{ \ j}=
g^{ik}f_{kj}-\frac{1}{2n^2} n^i n^k f_{kj} \ . \nonumber \\
\end{eqnarray}
The crucial point for the existence of four constraints that
generate the diagonal diffeomorphism  is to show that the potential
$\mV$ does not depend on $\bN$ and $\bN^i$. To do this we will argue
that the matrix $H^\mu_{ \ \nu}$ is similar to the matrix $\bA^\mu_{
\ \nu}$ defined as
\begin{equation}
\bA=
\left(\begin{array}{cc} a & v_j \\
w^i & a^i_{ \ j} \\ \end{array}\right) \ .
\end{equation}
Our arguments are as follows. It is known that matrices are
characterized by their characteristic polynomials. The
characteristic polynomial of an $n\times n$ matrix $\mathbf{X}$ is
given by
\begin{equation}
p(\lambda)=\det (\bX-\lambda \mathbf{I})=
(-1)^n[\lambda^n+c_1\lambda^{n-1}+c_2\lambda^{n-2}+\dots +
c_{n-1}\lambda+c_n] \ ,
\end{equation}
where $c_1,c_2,\dots,c_n$ are expressed using the powers of the
traces of $\mathbf{X}$. Explicitly, if we introduce the notation
\begin{equation}
t_k= \tr (\mathbf{X}^k)
\end{equation}
we find that in case of $4\times 4$ matrix  the characteristic
polynomial is determined by following coefficients
\begin{eqnarray}
c_1&=&-t_1 \ , \nonumber \\
c_2&=&\frac{1}{2}(t_1^2-t_2) \ , \nonumber \\
c_3&=&-\frac{1}{6}t^3_1+\frac{1}{2}t_1t_2-\frac{1}{3}t_3 \ ,
\nonumber
\\
c_4&=&\frac{1}{24}t_1^4-\frac{1}{4}t_1^2 t_2+\frac{1}{3}t_1t_3+
\frac{1}{8}t_2^2-\frac{1}{4}t_4 \ .  \nonumber \\
\end{eqnarray}
Let us now determine $t_k$ for the matrix $H^\mu_{ \ \nu}$
\begin{eqnarray}\label{ti}
t_1&=&\tr H=H^\mu_{ \ \mu}=a+a^i_{ \ i}=\bA^\mu_{ \ \mu} \ ,
\nonumber
\\
t_2&=&\tr H^2=H^\mu_{ \ \nu}H^\nu_{ \ \mu}=a^2+v_iw^i+a^i_{ \
j}a^j_{
\ i}=\bA^\mu_{ \ \nu} \bA^\nu_{ \ \mu} \ , \nonumber \\
t_3&=&\tr H^3= H^\mu_{ \ \nu}H^\nu_{ \ \rho} H^\rho_{ \ \mu}=
a^3+3v^iw_i a+3v_ia^i_{ \ j}w^j+a^i_{ \ j}a^j_{ \ k}a^k_{ \ i}=
\nonumber \\
&=& \bA^\mu_{ \ \nu}
\bA^\nu_{ \ \rho}\bA^\rho_{\  \mu} \ , \nonumber \\
t_4&=&\tr H^4=H^\mu_{ \ \nu}H^\nu_{ \ \rho} H^\rho_{ \
\sigma}H^\sigma_{ \ \mu}= \nonumber \\
&=& a^4+4a^2v_iw^i+2(v_iw^i)^2+ 4a v_ia^i_{ \ j}w^j+ 4v_i a^i_{ \
j}a^j_{ \ k}w^k+ a^i_{ \ j} a^j_{ \ k}a^k_{ \ l}a^l_{ \ i}=\nonumber
\\
&=&
 \bA^\mu_{ \ \nu}
\bA^\nu_{ \ \rho}\bA^\rho_{\ \sigma}\bA^\sigma_{ \ \mu} \ . \nonumber \\
\nonumber \\
\end{eqnarray}
We see that the characteristic polynomials of $H^\mu_{ \ \nu}$ and
$\bA^\mu_{\  \nu}$ are the same which also implies that
corresponding minimal polynomials are the same. Further, by
presumption $H^\mu_{ \ \nu}$ is diagonalizable which means that the
minimal polynomial is a product of distinct linear factors. However
since the minimal polynomial of $H^\mu_{ \ \nu}$ is the same as the
minimal polynomial of $\bA^\mu_{ \ \nu}$ we immediately find that
$\bA^\mu_{ \ \nu}$ is diagonalizable to the same diagonal matrix
which also implies that $H^\mu_{ \ \nu}$ and $\bA^\mu_{ \ \nu}$ are
similar.
 In other words there exist the matrix $T^\mu_{ \ \nu}$ so that
\begin{equation}
H^\mu_{ \ \nu}=T^\mu_{ \ \rho}\bA^\rho_{ \ \sigma}(T^{-1})^\sigma_{
\ \nu} \ .
\end{equation}
Then we  find
\begin{equation}\label{mVHA}
\mV(H)=\mV(\bA) \
\end{equation}
since by definition the potential is given as the traces of various
powers of $H^\mu_{ \ \nu}$. Now the equation (\ref{mVHA}) is the
desired result which shows that $\mV$ does not depend on  $\bN$ and
$\bN^i$ and hence they appear in the action linearly. Using these
calculations we find the Hamiltonian in the form
 \begin{equation}
H=\int d^3\bx (\bN\bmR+\bN^i\bmR_i) \ ,
\end{equation}
where
\begin{eqnarray}
\bmR&=&n\mR_0^{(g)}+\frac{1}{n}\mR_0^{(f)}+
\frac{1}{2}n^i\mR_i^{(g)}
-\frac{1}{2}n^i\mR_i^{(f)}+\nonumber \\
&+&\mu^2 g^{1/4}f^{1/4} \mV(\bA) \  , \quad   \bmR_i=
\mR_i^{(g)}+\mR_i^{(f)} \ . \nonumber \\
\end{eqnarray}
Now we proceed to the analysis of the requirement of the
preservation of the primary constraints (\ref{Pbnprim})
\begin{eqnarray}
\partial_t P_{\bN}&=&\pb{P_{\bN},H}=-\bmR
\approx 0 \ , \nonumber \\
\partial_t P_i&=&\pb{P_i,H}=-\bmR_i \approx 0 \ , \nonumber \\
\partial_t p_n&=&\pb{p_n,H}=-\mR_0^{(g)}+
\frac{1}{n^2}\mR_0^{(f)}-\mu^2 g^{1/4}f^{1/4}\frac{\delta
\mV}{\delta n}\equiv \mG_n\approx 0 \ , \nonumber \\
\partial_t p_i&=&\pb{p_i,H}=-\frac{1}{2}\mR_i^{(g)}+
\frac{1}{2}\mR_i^{(f)}-\mu^2 g^{1/4}f^{1/4}\frac{\delta
\mV}{\delta n^i}\equiv \mG_i\approx 0 \ . \nonumber \\
\end{eqnarray}
As a result we have following total Hamiltonian
\begin{eqnarray}\label{HT}
H_T=\int d^3\bx ( \bN\bmR+\bN^i\bmR_i+V_NP_{\bN}+V^i P_i +v_n
p_n+v^i p_i +u^n\mG_n+u^i\mG_i) \ ,
\nonumber \\
\end{eqnarray}
where $V_N,V^i,v_n,v^i,u^n,u^i$ are Lagrange multipliers
corresponding to the constraints
$P_{\bN}\approx 0 , P_i\approx 0 ,  p_n\approx 0 ,
p_i\approx 0, \mG_n\approx 0 , \mG_i\approx 0$.
In the next section we determine the algebra of constraints
$\bmR,\bmR_i$.  
\section{Algebra of Constraints $\bmR,\bmR_i$}\label{third}
For the consistency of the theory it is important to show that the
constraints $\bmR$ and $\bmR_i$ are the first class constraints.
 To proceed  it is useful to introduce the smeared form of
 the constraint $\bmR$
\begin{equation}\label{defsmearconsH}
\bT_T(N)=\int d^3\bx N(\bx)\bmR(\bx) \ .
\end{equation}
Instead of the  constraint $\bmR_i$ we introduce the constraint
$\tmR_i$ that is given as an extension of the constraints $\bmR_i$
by appropriate combinations of the primary constraints $p_n,p_i$
\begin{equation}
\tilde{\mR}_i=\mR_i^{(g)}+\mR_i^{(f)}+\partial_inp_n+\partial_in^jp_j+
\partial_j (n^j p_i) \
\end{equation}
with equivalent smeared form
\begin{equation}\label{defsmearconsHs}
\bT_S(N^i)=\int d^3\bx N^i\tilde{\mR}_i \ .
\end{equation}
Then using (\ref{canpb}) we determine the Poisson brackets between
$\bT_S(N^i)$ and the canonical variables
\begin{eqnarray}\label{bTScan}
\pb{\bT_S(N^i),g_{ij}}&=&-N^k\partial_k g_{ij}-\partial_i N^kg_{kj}-
g_{ik}\partial_jN^k \ ,  \nonumber \\
\pb{\bT_S(N^i),\pi^{ij}}&=& -\partial_k(N^k\pi^{ij})
+\partial_kN^i\pi^{kj}+\pi^{ik}\partial_k N^j \ , \nonumber \\
\pb{\bT_S(N^i),f_{ij}}&=& -N^k\partial_k f_{ij}-
\partial_iN^kf_{kj}-f_{ik}\partial_jN^k \ , \nonumber \\
\pb{\bT_S(N^i),\rho^{ij}}&=&-\partial_k (N^k\rho^{ij}) +\partial_k
N^i\rho^{kj}+\rho^{ik}\partial_k N^j \ , \nonumber \\
\pb{\bT_S(N^i),n}&=&-N^i\partial_i n \ , \nonumber \\
\pb{\bT_S(N^i),\pi_n}&=&-\partial_i (N^i\pi_n) \ , \nonumber \\
\pb{\bT_S(N^i),n^i}&=&-N^k\partial_k n^i+\partial_j N^in^j \ ,
\nonumber \\
\pb{\bT_S(N^i),\pi_i}&=&-\partial_k (N^k \pi_i)-\partial_i N^k\pi_k
\ .
\nonumber \\
\end{eqnarray}
With the help of  these results we  easily find how various
components $\bA^\mu_{ \ \nu}$ transform under spatial diffeomorphism
\begin{eqnarray}\label{btSa}
\pb{\bT_S(N^i),a}&=&-N^i\partial_i a \ ,  \nonumber \\
\pb{\bT_S(N^i),v_i}&=&-N^j\partial_jv_i-\partial_iN^kv_k \ ,
\nonumber
\\
\pb{\bT_S(N^i),w^j}&=&-N^k\partial_k w^j+w^k\partial_k N^j \ ,
\nonumber \\
\pb{\bT_S(N^i),a^i_{ \ j}}&=&-N^k\partial_k a^i_{ \ j}+
\partial_k N^i a^k_{ \ j}-a^i_{ \ k}\partial_j N^k \ . \nonumber \\
\end{eqnarray}
and also
\begin{equation}\label{pbbTSS}
\pb{\bT_S(N^i),\bT_S(M^j)}= \bT_S((N^j\partial_j M^i-M^j\partial_j
N^i)) \ .
\end{equation}
This result suggests that $\bT_S(N^i)$ is the generator of the
spatial diffeomorphism.

For further purposes it is convenient to introduce
 the smeared forms of the constraints $\mR_0^{(f),(g)}$
and $\mR_i^{(f),(g)}$
\begin{eqnarray}
 \bT_T^g(N)&=&\int d^3\bx N(\bx)\mR_0^{(g)}(\bx) \ , \quad
  \bT_T^f(N)= \int
d^3\bx N(\bx)\mR_0^{(f)}(\bx) \ , \nonumber \\
 \bT_S^g(N^i)&=&\int
d^3\bx N^i(\bx)\mR_i^{(g)}(\bx) \ , \quad
 \bT_S^f(N^i)=\int d^3\bx N^i(\bx)\mR_i^{(f)}(\bx) \ . \nonumber \\
\end{eqnarray}
It is well known that these smeared constraints have following
non-zero Poisson brackets \footnote{See, for example
\cite{Hojman:1976vp}.}
\begin{eqnarray}\label{pbGR}
\pb{\bT^g_T(N),\bT^g_T(M)}&=& \bT^g_S((N\partial_iM-M\partial_i
N)g^{ij}) \
, \nonumber \\
\pb{\bT^f_T(N),\bT^f_T(M)}&=& \bT^f_S((N\partial_iM-M\partial_i
N)f^{ij}) \
, \nonumber \\
\pb{\bT_S^g(N^i),\bT_T^g(M)}&=&\bT_T^g(N^i\partial_i M) \ , \nonumber \\
\pb{\bT_S^f(N^i),\bT_T^f(M)}&=&\bT_T^f(N^i\partial_i M) \ ,
\nonumber
\\
\pb{\bT_S^g(N^i),\bT_S^g(M^j)}&=&\bT^g_S((N^j\partial_j
M^i-M^j\partial_j N^i)) \ , \nonumber \\
\pb{\bT_S^f(N^i),\bT_S^f(M^j)}&=&\bT^f_S((N^j\partial_j
M^i-M^j\partial_j N^i)) \ . \nonumber \\
\end{eqnarray}
 As the next step we determine the Poisson bracket
between $\bT_S(N^i)$ and $\bT_T(N)$. We firstly determine following
Poisson bracket
\begin{eqnarray}\label{bTSNV1}
\pb{\bT_S(N),g^{1/4}f^{1/4}\mV}&=&
-N^k\partial_k[g^{1/4}f^{1/4}\mV]-\partial_k N^k g^{1/4}f^{1/4}\mV+
\nonumber \\
&+&g^{1/4}f^{1/4}\left[ \frac{\delta \mV}{\delta n^i}\partial_j
N^in^j- 2\frac{\delta \mV}{\delta f_{kl}}\partial_k N^m f_{ml}+
2\frac{\delta \mV}{\delta g^{kl}}\partial_m N^k g^{ml}\right] \ . \nonumber \\
\nonumber \\
\end{eqnarray}
On the other hand we know that $\mV$ depends on $\bA^\mu_{ \ \nu}$
through the trace. As a result we find that $\mV$ depends on the
eigenvalues $\lambda_i$ that are solutions of the characteristic
polynomials. Further, we know that the coefficients of this
polynomial are functions of $t_1,\dots,t_4$. As a result if we show
that $t_i$ transform as scalars under spatial diffeomorphism we find
that $\lambda_i$ and consequently $\mV$ transform as the scalar as
well. In fact, using the explicit form of $t_i$ given in (\ref{ti})
and also using the Poisson brackets (\ref{btSa}) we easily find that
$t_i$ transform as scalars under spatial diffeomorphism. These
results imply following Poisson bracket
\begin{equation}\label{bTSNV2}
\pb{\bT_S(N),g^{1/4}f^{1/4}\mV}= -N^k\partial_k[g^{1/4}f^{1/4}\mV]
-\partial_k N^k g^{1/4}f^{1/4}\mV \ .
\end{equation}
Finally using the Poisson brackets
(\ref{pbGR}) we find
\begin{eqnarray}
\pb{\bT_S(N^i),\bT_T(M)}&=&\bT_T(N^i\partial_iM)  \ , \nonumber \\
\pb{\bT_S(N^i),\bG_n(M)}&=&\bG_n(N^i\partial_iM) \ . \nonumber \\
\end{eqnarray}
In case of $\mG_i$ the situation is more complicated since we have
to determine the variation of $t_i$ with respect  to $n^j$ and then
its Poisson bracket with  $\bT_S(N^i)$. For example, in case $t_1$
and $t_2$ we obtain
\begin{eqnarray}
\frac{\delta t_1}{\delta n^i}&=&-2\frac{f_{ij}n^j}{n^2} \ ,
\nonumber
\\
\frac{\delta t_2}{\delta
n^i}&=&-\frac{a}{n^2}f_{ij}n^j+\frac{1}{n^2}f_{ij}w^j
-\frac{v_i}{2}a-\frac{1}{2}f_{ik}g^{kj}v_j \ .  \nonumber \\
\end{eqnarray}
In the same way we can calculate $\frac{\delta t_3}{\delta n^i}$ and
$\frac{\delta t_4}{\delta n^i}$  and then we find
\begin{equation}
\pb{\bT_S(N^i),\frac{\delta t_j}{\delta n^i}}= -N^k\partial_k
\left[\frac{\delta t_j}{\delta n^i}\right]-\frac{\delta t_j}{\delta
n^k}
\partial_i N^k \ .
\end{equation}
With the help of these results we easily obtain
\begin{eqnarray}\label{pbbTSmVn}
\pb{\bT_S(N^i), \frac{\delta \mV}{\delta n^i}}= \frac{\delta
\mV}{\delta t_j} \pb{\bT_S(N^i),\frac{\delta t_j}{\delta n^i}}=
-N^k\partial_k\left[\frac{\delta \mV}{\delta n^i}\right]-
\frac{\delta \mV}{\delta n^k}\partial_i N^k \ .  \nonumber \\
\end{eqnarray}
Finally using (\ref{pbGR}) and (\ref{pbbTSmVn})  we find
\begin{equation}
\pb{\bT_S(N^i),\mG_i}=-N^k\partial_k\mG_i-\mG_j\partial_i N^j \ .
\end{equation}
Collecting all these results we find that $\tmR_i$ are the first
class constraints whose smeared forms correspond to the generator of
the diagonal spatial diffeomorphism.

On the other hand more interesting is to determine the Poisson
bracket between smeared forms of the Hamiltonian constrains
(\ref{defsmearconsH}). Following the same calculations as in
\cite{Kluson:2012ps} we find
\begin{eqnarray}\label{pbbTT}
& &\pb{\bT_T(N),\bT_T(M)}
 =\bT_S((N\partial_i M-M\partial_i N)n^2 g^{ij})+
\bT_S((N\partial_i M-M\partial_i N)\frac{1}{n^2}f^{ij})
 -\nonumber \\
 &-&\bG_S((N\partial_i M-M\partial_i N)n^2 g^{ij})
-\bG_T((N\partial_iM-M\partial_iN)n^i)
+\nonumber \\
&+&\bG_S((N\partial_i M-M\partial_iN)\frac{1}{n^2}f^{ij}) +\int
d^3\bx
(N\partial_i M-M\partial_i N)\Sigma^i[\mV] \ , \nonumber \\
\end{eqnarray}
where we defined the smeared forms of the constraints $\mG_i$ and
$\mG_n$
\begin{equation}
\bG_T(N)=\int d^3\bx N(\bx) \mG_n(\bx)
 \ , \quad  \bG_S(N^i)= \int d^3\bx
N^i(\bx)\mG_i(\bx) \ ,
\end{equation}
and where $\Sigma^i[\mV]$ is defined as
\begin{eqnarray}
\Sigma^i[\mV]=\mu  g^{1/4}f^{1/4}\left[-n^2g^{ij}
 \frac{\delta \mV}{\delta n^j}
 +f^{ij}\frac{\delta \mV}{\delta n^j}\frac{1}{n^2}
- \frac{\delta \mV}{\delta g^{kj}}n^kg^{ij}- \frac{\delta
\mV}{\delta f_{ij}} n^kf_{kj} -\frac{1}{2}n^in\frac{\delta
\mV}{\delta n} \right] \ .  \nonumber \\
\end{eqnarray}
We see that the Poisson bracket between the smeared forms of the
Hamiltonian constraints (\ref{pbbTT}) vanish on the constraint
surface on condition that $\Sigma^i[\mV]$ is zero.
In order to calculate $\Sigma^i[\mV]$ we proceed in the similar way
as in case when we calculated the Poisson bracket between
$\bT_S(N^i)$ and $\mV$. Explicitly, we know that $\bA^\mu_{ \ \nu}$
is similar to diagonal matrix with eigenvalues
$\lambda_1,\lambda_2,\lambda_3,\lambda_4$. In other words we have
\begin{equation}
\mV(\bA)=\mV(\lambda_i) \ .
\end{equation}
On the other hand we know that $\lambda_i$ are solutions of the
characteristic polynomial when the coefficients are functions of
$t_1,\dots,t_4$ so that
\begin{equation}
\lambda_i=\lambda_i(t_1,\dots,t_4) \ .
\end{equation}
Then if we show that
\begin{equation}
\Sigma^i[t_j]=0 \ , j=1,2,3,4
\end{equation}
we find that
\begin{equation}
\Sigma^i[\lambda_j]=0
\end{equation}
and consequently
\begin{equation}
\Sigma^i [\mV(\lambda)]= \Sigma^i [\mV(\bA)]=0 \ .
\end{equation}
We have already found in \cite{Kluson:2012ps} that $\Sigma^i
[t_1]=\Sigma^i[t_2]=0$. The case of $t_3$ and $t_4$ is more
intricate. For example, the explicit form of $t_3$ is equal to
\begin{eqnarray}
t_3&=&\frac{1}{n^{12}}-\frac{3}{n^{10}}n^if_{ij}n^j+\frac{3}{n^8}(n^if_{ij}n^j)^2
-\frac{1}{n^6}(n^if_{ij}n^j)^3
-\nonumber \\
&-&\frac{3}{n^6} n^if_{ik}g^{kl}f_{lm}n^m+ \frac{3}{n^4}
(n^if_{ik}g^{kl}f_{lm}n^m)(n^r f_{rs}n^s)-\nonumber \\
&-& \frac{3}{n^2} n^mf_{mi}g^{ip}f_{pj}g^{jr}f_{rk}n^k +
g^{im}f_{mj}g^{jk}f_{kl}g^{lp}f_{pi} \ .  \nonumber \\
\end{eqnarray}
Then using the explicit form of $\Sigma^i$ and after some
calculations we find
\begin{equation}
\Sigma^i[t_3]=0 \ .
\end{equation}
In the same way  we find $\Sigma^i[t_4]=0$. In summary  we have
\begin{equation}
\Sigma^i[t_j]=0 \ , \  \mathrm{for} \  j=1,2,3,4 \ .
\end{equation}
With the help of this result we obtain
 the fundamental result that
the Poisson bracket between Hamiltonian constraint (\ref{pbbTT})
vanishes on the constraint surface. This result together with
(\ref{pbbTSS}) and (\ref{pbbTT}) shows that $\bT_T(N)$ and
$\bT_S(N^i)$ are the first class constraints that are generators of
the diagonal  diffeomorphism.

\section{Time Evolution of Remaining
Constraints}\label{fourth}
 In this section we analyze the time
evolution of the constraints $p_n,p_i,\mG_n,\mG_i$.  Note that the
total Hamiltonian is given in (\ref{HT}).
For further purposes we calculate following Poisson brackets
\begin{eqnarray}\label{PnmG}
\pb{p_n(\bx),\mG_i(\by)}&=&
 \mu^2 \frac{\delta^2 \mV}{\delta n^i
\delta n}(\bx)\delta(\bx-\by)\equiv \triangle_{ni}(\bx)
\delta(\bx-\by) \ , \nonumber \\
 \pb{p_n(\bx),\tmG_n(\by)}&=&
-\frac{1}{n}\mG_n(\bx)\delta(\bx-\by)+\nonumber \\
&+&
 n(\bx)\left[ \mR_0^{(g)} +\frac{1}{n^2}\mR_0^{(f)}+
\mu^2 g^{1/4}f^{1/4}\frac{\delta \mV}{\delta n}+
 \mu^2g^{1/4}f^{1/4}n\frac{\delta^2 \mV}{\delta n(\bx)\delta
 n(\by)}\right]\delta(\bx-\by)\approx\nonumber \\
&\approx &n(\bx)\left[\frac{1}{n}\bmR
+\frac{n^i}{n}\mG_i+\frac{n^i}{n}\mu^2 g^{1/4}f^{1/4}\frac{\delta
\mV}{\delta n^i}
 -\frac{\mu^2}{n}g^{1/4}f^{1/4}\mV+\right.\nonumber \\
 &+& \left.\mu^2 g^{1/4}f^{1/4}\frac{\delta \mV}{\delta n}+
 \mu^2g^{1/4}f^{1/4}n\frac{\delta^2 \mV}{\delta^2 n}\right]\delta(\bx-\by)\approx \nonumber \\
&\approx & \mu^2 g^{1/4}f^{1/4}\left[n^i\frac{\delta \mV}{\delta
n^i} -\mV
 + n\frac{\delta \mV}{\delta n}+
 n^2\frac{\delta^2 \mV}{\delta^2 n}\right]  \delta(\bx-\by)\equiv \triangle_{nn}(\bx-\by) \ ,  \nonumber \\
 \pb{p_i(\bx),\mG_n(\by)}&=&\mu^2g^{1/4}f^{1/4}\frac{\delta^2\mV}{\delta
 n^i\delta n}(\bx)\delta(\bx-\by)\equiv
 \triangle_{in}(\bx)\delta(\bx-\by) \ , \nonumber \\
 \end{eqnarray}
We introduce following common notation $p_a\equiv (p_n,p_i)$ and $
\mG_a\equiv (\mG_n,\mG_i)$. Then the requirement of the preservation
of the primary constraints $p_a$ takes the form
\begin{eqnarray}\label{timePn}
\partial_t p_a&=&\pb{p_a,H_T}=\mG_n+
\int d^3\bx u^b(\bx)\pb{p_a,\mG_b(\bx)} \approx
\nonumber \\
&\approx& \int d^3\bx u^b(\bx)\pb{p_a,\mG_b(\bx)}=\triangle_{ab}u^b
=0 \ .
\nonumber \\
\end{eqnarray}
We have four equations for four unknown $u^a$. We have to
distinguish two cases. The case when $\det \triangle_{ab}\neq 0$ and
the case when $\det \triangle_{ab}=0$.
\subsection{The case $\det \triangle_{ab}\neq 0$}
This situation is particular simple and we believe that this is in
fact the situation in all  examples of the bimetric theories of
gravity that are invariant under diagonal diffeomorphism. In this
case we find that the solution of the equations (\ref{timePn}) is
$u^b=0$.
%
Then we can proceed to the analysis of the preservation of the
constraints $\mG_a$
\begin{eqnarray}
\partial_t\mG_a=\pb{\mG_a,H_T}\approx
\int d^3\bx \left(N(\bx)\pb{\mG_a,\bR(\bx)}+
v^b\pb{\mG_a,p_b(\bx)}\right)=
\nonumber \\
=\int d^3\bx N(\bx)\pb{\mG_a,\bR(\bx)}- \triangle_{ab}v^b=0
\nonumber \\
\end{eqnarray}
that again using the fact that $\det \triangle_{ab}\neq 0$ can be
solved for $v^a$ as functions of the canonical variables and $N$.
 In other words we
find that $\mG_a,p_a$ are the second class constraints that can be
solved for $n,n^i$ at least in principle. Then four first class
constraints $P_{\bN},P_i$ can be gauge fixed and  hence we can
eliminate $P_{\bN},P_i$ and $\bN,\bN^i$ as dynamical variables. Then
remaining four first class constraints $\bmR,\tmR_i$ can be again
gauge fixed so that we can eliminate eight degrees of freedom from
$24$ degrees of freedom $g_{ij},\pi^{ij},f_{ij},\rho^{ij}$ so that
we have $16$ physical degrees of freedom where $4$ correspond to the
massless graviton, $10$ to the massive one and $2$ corresponding to
the scalar mode at least at the linearized level. Of course, this
Hamiltonian analysis cannot answer an important question which is
the identification of the physical metric that couples to the matter
fields, for recent detailed analysis, see \cite{Hassan:2012wr}.

To conclude this section we have to mention the form of the
symplectic structure on the reduced phase space spanned by
$z^\alpha\equiv(g_{ij},\pi^{ij},f_{ij},\rho^{ij})$. Due to the fact
that we have eight  second class constraints $p_n,p_i,\mG_n,\mG_i$
we should replace the Poisson brackets with corresponding Dirac
brackets.
 Explicitly, let us introduce the common notation for all second class constraints
$\chi_A\equiv (P_a,\mG_a)$. Then we find following matrix of the
Poisson brackets of the second class constraints $\chi_A$
\begin{equation}
\pb{\chi_A(\bx),\chi_A(\by)}\equiv \Theta_{AB}
=\left(\begin{array}{cc} 0 & \triangle(\bx)\delta(\bx-\by) \\
-\triangle(\by)\delta(\bx-\by) & \Omega(\bx,\by) \\
\end{array}\right) \ ,
\end{equation}
where  the matrix $\Omega_{ab}(\bx,\by)$ is defined as
\begin{equation}\label{defOmega}
\pb{\mG_a(\bx),\mG_b(\by)}=\Omega_{ab}(\bx,\by) \ .
\end{equation}
Since $\triangle$ is invertible matrix   we can find the matrix
inverse to $\Theta_{AB}$ in the form
\begin{equation}
(\Theta^{-1})^{AB}= \left(\begin{array}{cc}
\triangle^{-1}(\bx)\Omega(\bx,\by) \triangle^{-1}(\by) &
-\triangle^{-1}(\bx)\delta(\bx-\by) \\
\triangle^{-1}(\by)\delta(\bx-\by) & 0 \\ \end{array}\right)
\end{equation}
Following the standard analysis of the system with the second class
constraints we replace the Poisson bracket by corresponding Dirac
bracket defined as
\begin{eqnarray}
\pb{F,G}_D= \pb{F,G}- \int d^3\bx d^3\by \pb{F,\chi_A(\bx)}
(\Theta^{-1})^{AB}(\bx,\by) \pb{\chi_B(\by),G} \ .  \nonumber \\
\end{eqnarray}
Now we see that for the variables on the reduced phase space spanned
by $z^\alpha\equiv (g_{ij},\pi^{ij},f_{ij},\rho^{ij})$ the Dirac
bracket coincides with the Poisson bracket. Explicitly, we have
\begin{eqnarray}
& &\pb{z^\alpha(\bx),z^\beta(\by)}_D=
\pb{z^\alpha(\bx),z^\beta(\by)}-
\nonumber \\
&-& \int d^3\bz d^3\bz' \pb{z^\alpha(\bx),\chi_A(\bz)}
(\Theta^{-1})^{AB}(\bz,\bz') \pb{\chi_B(\bz'),z^\beta(\bx)}=
 \pb{z^\alpha(\bx),z^\beta(\by)}-
\nonumber \\
&-& \int d^3\bz d^3\bz' \sum_{A=4}^7\sum_{B=4}^7
\pb{z^\alpha(\bx),\chi_A(\bz)} (\Theta^{-1})^{AB}(\bz,\bz')
\pb{\chi_B(\bz'),z^\beta(\bx)}=\pb{z^\alpha(\bx),z^\beta(\by)}
\nonumber \\
\end{eqnarray}
due to the fact that $(\Theta^{-1})^{AB}=0$ for $A,B=3,\dots,7$ are
zero.
\subsection{The case $\det \triangle_{ab}=0$}
In this case it is possible to find the vector $u^a_0$ such that
\begin{equation}
\triangle_{ab}u^b_0=0 \ .
\end{equation}
Then we define new constraints $\tP,\tmG$ defined as
\begin{equation}
\tP=u^a_0 P_a \ , \tmG=u^a_0\mG_a \
\end{equation}
together with
\begin{equation}
\tP_a=P_a-\frac{1}{u_{0a} u^a_0}u_{0a}\tP \ , \quad \tmG_a=\mG_a
-\frac{1}{u_{a0}u^a_0}u_{0a}\tmG \ , \quad u_{0a}=\delta_{ab}u^b_0
\end{equation}
that by definition obey the conditions
\begin{equation}
u^a_0\tP_a=0 \ , \quad u^a_0\tmG_a=0 \
\end{equation}
which means that we have $6$ independent constraints.
 As a result we consider the total
Hamiltonian in the form
\begin{eqnarray}\label{Hamtot}
H_T=\int d^3\bx (N\tmR+N^i\tmR_i+V_{\bN}P_{\bN}+ V^iP_i+\tv \tP+
\tv^a\tP_a+ \tu \tmG+
\tu^a\tmG_a) \ , \nonumber \\
\end{eqnarray}
where
\begin{equation}\label{deftvau}
\tv^a=v^a-\frac{u^a_0(v^au_{0a})}{u^au_{0a}} \ , \quad \tu^a= u^a-
\frac{(u^a u_{0a})u^a_0}{u^a_0 u_{0a}} \ ,
\end{equation}
where  $v^a,u^a$ are arbitrary four dimensional vectors and we
clearly see that $\tv^a, \tu^a$ are orthogonal to $u^a_0$ so that
they have  $3$ independent components.

Now we proceed to the analysis of the time evolution of the
constraints $\tP,\tP_a,\tmG,\tmG_a$. Note that the constraint $\tP$
has vanishing Poisson bracket with $\tmG$ and $\tmG_a$ since
\begin{eqnarray}
\pb{\tP(\bx),\tmG(\by)}&=&u^a_0\triangle_{ab}u^b_0\delta(\bx-\by)=0
\
, \quad \nonumber \\
\pb{\tP(\bx),\tmG_a(\by)}&=& u^b_0\triangle_{ba}\delta(\bx-\by)=0 \
\nonumber \\
\end{eqnarray}
so that it is trivial to see that $\tP$ is preserved during the time
evolution of the system and that it is the first class constraint.
Let us now analyze the time evolution of the constraint $\tP_a$
\begin{eqnarray}\label{parttPa}
\partial \tP_a=\pb{\tP_a,H_T}\approx
\int d^3\bx ( \tu(\bx)\pb{\tP_a,\tmG(\bx)} + \tu^b(\bx)
\pb{\tP_a,\tmG_b(\bx)})\approx \triangle_{ab}  \tu^b =0\nonumber \\
\end{eqnarray}
using
\begin{equation}
\pb{\tP_a(\bx),\tmG(\by)}=\pb{P_a(\bx),\mG_b(\by)}u^b_0=
\triangle_{ab}u^b_0\delta(\bx-\by)=0 \ .
\end{equation}
Now due to the fact that $\tu^a$ is orthogonal to $u^a_0$ we find
that the only solution of (\ref{parttPa}) is given by $\tu^a=0$.

We proceed to the analysis of the time evolution of the constraint
$\tmG,\tmG_a$. We start with the constraints $\tmG_a$
\begin{eqnarray}
\partial_t \tmG_a=
\pb{\tmG_a,H_T}\approx \int d^3\bx
\left(N(\bx)\pb{\tmG_a,\bmR(\bx)}+
\tu(\bx)\pb{\tmG_a,\tmG(\bx)}\right)+ \tv^b\triangle_{ba}=0 \nonumber \\
\end{eqnarray}
that again using the definition of $\tv^b$ given (\ref{deftvau}) can
be solved for $\tv^a$ as function of the canonical variables and $N$
and $\tu$. This result implies that $\tP_a,\tmG_b$ are the second
class constraints.

Now we come the most intricate analysis which is the time
preservation of the constraint $\tmG$
\begin{eqnarray}\label{parttmG}
\partial_t\tmG=\pb{\tmG,H_T}
\approx \int d^3 \bx \left(N(\bx)\pb{\tmG,\bmR(\bx)}+
\tu(\bx)\pb{\tmG,\tmG(\bx)}\right)=0  \ . \nonumber \\
\end{eqnarray}
Note that generally $\pb{\tmG(\bx),\tmG(\by)} \neq 0$. However  we
see that we have to analyze the time evolution of the constraint
$\bmR$ as well \footnote{Note that  this was not necessary in case
when $\det \triangle_{ab}\neq 0$ since in this case we found that
$u^a=0$ and hence the constraint $\tmR$ is trivially preserved.} so
that
\begin{eqnarray}\label{presermR}
\partial_t\bmR=\pb{\bmR,H_T}
\approx \int d^3\bx \tu(\bx)\pb{\bmR,\tmG(\bx)}=0 \  . \nonumber \\
\end{eqnarray}
To proceed further we have to determine the Poisson bracket between
$\bmR$ and $\tmG$. This is very complicated due to the complex form
of these constraints. As an example we determine the Poisson bracket
between $\bT_T(N)$ and $\bG_T(M)$
\begin{eqnarray}\label{bTbG}
\pb{\bT_T(N),\bG_T(M)}
&=&-\bT^g_S(n(N\partial_i
M-M\partial_iN)g^{ij}-MN\partial_ing^{ij})+
\nonumber \\
&+& \bT^f_S(n^{-3}(N\partial_iM-M\partial_i N)f^{ij}- n^{-4}
\partial_i n NM f^{ij})-\nonumber \\
&-&\frac{1}{2}\bT_T^g(n^iN\partial_iM)- \frac{1}{2}
\bT_T^f(n^iN\partial_i (n^{-2}M))+
\nonumber \\
&+&2\mu^2\int d^3\bx MN\left(\frac{1}{M_g^2
\sqrt{g}}n\pi^{kl}\mG_{ijkl}G^{kl}-
\frac{1}{M_f^2n\sqrt{f}}\rho^{kl}\mG_{klij}F^{ij}\right)+ \nonumber \\
&+&\frac{1}{2}\mu^2\int d^3\bx G^{ij}_n\left(Nn^k\partial_k g_{ij}+
\partial_i (Nn^k)g_{kj}+g_{ik}\partial_j(Nn^k)\right)-
\nonumber \\
&-&\frac{1}{2}\mu^2\int d^3\bx F^{ij}_n\left(Nn^k\partial_k f_{ij}+
\partial_i (Nn^k)f_{kj}+f_{ik}\partial_j(Nn^k)\right)-
\nonumber \\
&-&2\mu^2\int d^3\bx MN\left(\frac{1}{M_g^2
\sqrt{g}}n\pi^{kl}\mG_{ijkl}G^{kl}-
\frac{1}{M_f^2n\sqrt{f}}\rho^{kl}\mG_{klij}F^{ij}\right) \ ,  \nonumber \\
\end{eqnarray}
where
\begin{eqnarray}
G^{kl}&=&\frac{\delta (g^{1/4}f^{1/4} \mV)} {\delta g_{kl}} \ ,
\quad F^{kl}= \frac{\delta (g^{1/4}f^{1/4} \mV)}{\delta f_{kl}} \ ,
\nonumber \\
G^{kl}_n&=&\frac{\delta (g^{1/4}f^{1/4}\frac{\delta \mV}{\delta n})}
{\delta g_{kl}} \ , \quad  F^{kl}_n= \frac{\delta
(g^{1/4}f^{1/4}\frac{\delta \mV}{\delta n})}{\delta f_{kl}} \ .
\nonumber \\
\end{eqnarray}
In the same way we  calculate the Poisson bracket between $\bT(N)$
and $\bG_S(N^i)$ and we  derive similar result. Then in  principle
we could express the  Poisson bracket  between $\bT(N)$ and $\int
d^3\bx M(\bx)\tmG(\bx)$ in the form
\begin{eqnarray}\label{pbbCbDs}
& &\pb{\bT_T(N),\int d^3\bx M(\bx)\tmG(\bx)}= \nonumber \\
&=& \int d^3\bz ( N(\bz)M(\bz) \bF(\bz)
 +\partial_{z^i} N(\bz)\bV^i(\bz)M(\bz)+
N(\bz)\partial_{z^i} M(\bz) \bW^i(\bz)) \ ,
\nonumber \\
\end{eqnarray}
where the explicit form of $\bF,\bV^i,\bW^i$ can be derived from the
Poisson brackets between $\bT(N)$ and $\bG_T(N),\bG_S(N^i)$ and from
the known vector $u_0^a$.
 For our purposes it is useful the local form of the
Poisson bracket (\ref{pbbCbDs})
\begin{equation}\label{locbmRmG}
\pb{\bmR(\bx),\tmG(\by)}= \delta(\bx-\by)\bF(\bx)+
 \frac{\partial}{\partial y^i}
 \delta(\bx-\by)\bV^i(\by)+\frac{\partial}{\partial x^i}
 \delta(\bx-\by)\bW^i(\bx) \ .
\end{equation}
With the help of this result we find that (\ref{presermR}) takes the
form
\begin{eqnarray}\label{presermR2}
\partial_t\bmR=
\tu (\bF-\partial_i \bV^i)+\frac{\partial \tu}{\partial x^i}
(\bW^i-\bV^i)=0  . \nonumber \\
\end{eqnarray}
In order to find the meaning of this result we can argue
 in the same way as in \cite{Kluson:2013cy} where more details can be found.
 Briefly, as we can deduce  from (\ref{bTbG}) we have that $\bW^i\neq \bV^i$ so that
we should interpret the equation (\ref{presermR2}) as the
differential equation for $\tu(\bx)$ and this equation  could be
solved for $\tu(\bx)$ at least in principle. Further, using
(\ref{locbmRmG}) we find that  (\ref{parttmG}) becomes differential
equation for $N(\bx)$
\begin{eqnarray}
\partial_t\mG_n&=&
N(-\bF+\partial_i W^i)+\frac{\partial N}{\partial y^i}
(\bW^i-\bV^i)+ \nonumber \\
&+&\int d^3\by \tu(\by)\pb{\mG_n(\bx),\mG_n(\by)}=0 \ .  \nonumber \\
\end{eqnarray}
This equation can be now solved for $N(\bx)$ as function of the
canonical variables since the Lagrange multiplier $\tu$ has been
already determined by the equation (\ref{presermR}). These results
imply that $\bmR$ and $\tmG$ are the second class constraints. On
the other hand we could  say that  by consistency the theory should
have the first class  Hamiltonian constraint as a consequence of the
fact that the theory is invariant under diagonal diffeomorphism. In
other words we are tempting to say that in bimetric theories that
are invariant under diagonal diffeomorphism the case when  $\det
\triangle_{ab}=0$ cannot occur \footnote{The situation is different
in the miracle case when $\bW^i=\bV^i$. In this case it is more
natural to interpret $ \mG^{(3)}\equiv\bF-\partial_i\bV^i=0$ as the
new constraint rather than impose the condition $\tu=0$
\cite{Kluson:2013cy}. Then of course the equation (\ref{parttmG})
vanishes on the constraint surface $\mG^{(3)}\approx 0$ and the
requirement of the preservation of the constraint $\mG^{(3)}\approx
0$ leads to the emergence of the new constraint $\mG^{(4)}\approx 0$
on condition that the Poisson bracket between $\mG^{(3)}$ and $\bmR$
does not contain the derivative of the delta function. However due
to the complex form of the constraints $\bmR$ and $\mG_a$ and
corresponding Poisson brackets we believe that such a case is highly
improbable.}.

\subsection{Square root potential} Recently it was shown that the
non-linear massive gravity with the square root form of the
potential  is free from ghosts.
 The extension to the
bimetric gravity was performed in \cite{Hassan:2011zd} where it was
also argued that resulting theory should be ghost free. This fact
was questioned in \cite{Kluson:2013cy} where the Hamiltonian
analysis of the bimetric gravity introduced in \cite{Hassan:2011zd}
was performed. We argued that the theory with redefined shift
function   lost the manifest diffeomorphism invariance and so that
it difficult to find the generator of the diagonal diffeomorphism.
We  also argued that the number of the physical degrees of freedom
is $16$ so  that the scalar mode cannot be eliminated.

In order to resolve the apparent issue of  the impossibility to
identify the generators of the diagonal diffeomorphism in the
non-linear bimetric theory of gravity with redefined shift function
\cite{Hassan:2011zd} we can certainly use the analysis of the
general bimetric theory of gravity presented in this paper.
Obviously we can identify  four first class constraints $\bmR,\tmR_i
$ corresponding to the diagonal diffeomorphism. On the other hand we
can ask the question whether there is  possibility to identify an
additional constraint that could eliminate the scalar mode. To be
more concrete, let us consider following  square root potential
\begin{equation}
\mV=\tr\sqrt{H^\mu_{ \ \nu}}=\tr\sqrt{\bA^\mu_{ \ \nu}} \ .
\end{equation}
The problem is that it is not easy task to find the square root of
the matrix $\bA^\mu_{ \ \nu}$ since we were not able to find such
redefinition of the canonical fields as in
\cite{Hassan:2011tf,Hassan:2011hr} that could allow us to find the
explicit form of the square root of $\bA^\mu_{ \ \nu}$. For that
reason we try to find such a matrix in the following way. We
temporally write $n^i$ with the parameter $ \epsilon$ so that
$n^i\rightarrow \epsilon n^i$ when we take $\epsilon=1$ in the end.
Then we can write
\begin{equation}
\bA=\epsilon^0\bE^{(0)}+\epsilon \bE^{(1)}+ \epsilon^2\bE^{(2)}+
\epsilon^3\bE^{(3)} \ ,
\end{equation}
where
\begin{eqnarray}
\bE^{(0)}&=& \left(\begin{array}{cc} \frac{1}{n^4} & 0 \\
0 & g^{ik}f_{kj} \\ \end{array}\right) \ , \quad
\bE^{(1)}= \left(\begin{array}{cc}0 & \frac{n^kf_{kj}}{n^2} \\
-\frac{n^i}{2n^4} & 0 \\ \end{array}\right) \ ,
\nonumber \\
\bE^{(2)}&=&\left(\begin{array}{cc} -\frac{n^if_{ij}n^j}{2n^2} & 0 \\
0 & -\frac{n^in^kf_{kj}}{2n^2} \\ \end{array}\right) \ , \quad
\bE^{(3)}=
\left(\begin{array}{cc} 0 & 0 \\
\frac{1}{4n^2} n^i(n^mf_{mn}n^n) & 0 \\ \end{array}\right) \ .
\nonumber \\
\end{eqnarray}
We can solve $\sqrt{\bA}$ as the  expansion with respect to
$\epsilon$
\begin{equation}
\sqrt{\bA}=\sum_{n=0}^\infty \epsilon^n \bB^{(n)} \ .
\end{equation}
Taking square of given expression and comparing expressions of given
order in $\epsilon$ we find
\begin{eqnarray}\label{bEbB}
\bE^{(0)}&=&\bB^{(0)}\bB^{(0)} \ , \nonumber \\
\bE^{(1)}&=&\bB^{(0)}\bB^{(1)}+\bB^{(1)}\bB^{(0)} \ , \nonumber \\
\bE^{(2)}&=&\bB^{(0)}\bB^{(2)}+\bB^{(2)}\bB^{(0)}+\bB^{(1)}\bB^{(1)}
\
, \nonumber \\
\bE^{(3)}&=&\bB^{(0)}\bB^{(3)}+\bB^{(1)}\bB^{(2)}+\bB^{(2)}\bB^{(1)}+
\bB^{(3)}\bB^{(0)}  \ , \nonumber \\
0&=&\bB^{(1)}\bB^{(3)}+\bB^{(2)}\bB^{(2)}+\bB^{(3)}\bB^{(1)}+
\bB^{(0)}\bB^{(4)}+\bB^{(4)}\bB^{(0)} \ , \nonumber \\
&\vdots & \ . \nonumber \\
\end{eqnarray}
From the first equation  in (\ref{bEbB}) we find
\begin{equation}
\bB^{(0)}=\left(\begin{array}{cc}
\frac{1}{n^2} & 0 \\
0 & \gamma^i_{ \ j} \\ \end{array}\right) \ ,
\end{equation}
where
\begin{equation}
\gamma^i_{ \ j}=\sqrt{g^{ik}f_{kj}}  \ , \quad \gamma^i_{ \
k}\gamma^k_{ \ j}=g^{ik}f_{kj} \ .
\end{equation}
Further, the second equation in (\ref{bEbB}) can be solved as
\begin{equation}
\bB^{(1)}=\left(\begin{array}{cc}
0 & \frac{n^kf_{kl}}{n^2}(M^{-1})^l_j \\
(M^{-1})^i_{ \ k}(-\frac{n^k}{2n^4}-\frac{1}{2} g^{km} f_{mn}n^n )&
0 \\ \end{array}\right)
\end{equation}
where
\begin{equation}
M^i_{ \ k}=\frac{1}{n^2}\delta^i_{ \ k}+\gamma^i_{ \ k} \ .
\end{equation}
Repeating this procedure we could  determine all matrices
$\bB^{(n)}$ at least in principle. On the other hand we see that the
resulting expression contains all powers of $n$ and $n^i$. Then from
(\ref{PnmG}) we can deduce that $\triangle_{ab}\neq 0$ and hence
$p_a,\mG_a$ are the second class constraints without possibility to
eliminate additional scalar mode.
\section{Conclusion}\label{fifth}
Let us outline the results derived in this paper. We performed the
Hamiltonian analysis of bimetric theory of gravity with general form
of the potential. We found the first class constraints that are
generators of the diagonal diffeomorphism. Then we analyzed the
requirement of the preservation of remaining constraints during the
time development of the system. We show that there are two
possibilities. In the first case  when the determinant $\det
\triangle_{ab}\neq 0$ we found that there are eight second class
constraints. Unfortunately these second class constraints are not
sufficient for the elimination of the scalar  mode. On the other
hand we showed that in the second case $\det\triangle_{ab}=0$ we
find the first class constraint $\tP$. However then we argued that
in this case the Hamiltonian constraint $\bmR$ together with the
constraint $\tmG$ are the second class constraints. On the other
hand since we know that the theory is invariant under diagonal
diffeomorphism we suggests that the Hamiltonian constraint should be
the first class constraint and consequently the case when
$\det\triangle_{ab}=0$ should not occur. All these results could
also imply that the ghost mode is still presented in the non-linear
bimetric theory of gravity.

 As the extension of this work we suggest the Hamiltonian analysis
of the multimetric theories of gravity that were introduced in paper
\cite{Hinterbichler:2012cn} and we hope to perform such analysis
soon.

 \noindent {\bf
Acknowledgement:}
 This work   was
supported  by the Grant agency of the Czech republic under the grant
P201/12/G028. \vskip 5mm


\end{document}